\documentstyle[twoside,fps,fleqn,espcrc2]{article}
\input{psfig}
\newcommand\be{\begin{equation}}
\newcommand\ee{\end{equation}}
\newcommand\bea{\begin{eqnarray}}
\newcommand\eea{\end{eqnarray}}

\title{\Large\bf Calculation of Moments of Nucleon Structure
                 Functions\thanks{This work supported by  the U.S. 
                 Department of Energy (D.O.E.) under cooperative 
                 research agreement DE-FC02-94ER40818.}}

\author{R.C.~Brower, S.~Huang, J.W.~Negele, A.~Pochinsky and B.~Schreiber
        \address{Center for Theoretical Physics, LNS, Massachusetts
        Institute of Technology,\\ Cambridge MA 02139, USA}}

\begin{document}

\begin{abstract} 
Preliminary results are presented in our program to calculate low
moments of structure functions for the proton and neutron on a
$24^3\times32$ lattice at $\beta=6.2$. A comparison is made for a
variety of smeared nucleon sources and preliminary results for the
calculation of the nucleon tensor charge are presented.
\end{abstract}

\maketitle

\section{INTRODUCTION}

There are now extensive experimental results on nucleon structure
functions in deep inelastic scattering, providing a detailed empirical
knowledge of the distribution of quarks and gluons in the nucleon.
Thus a major theoretical challenge for lattice QCD is to understand
this data from first principles and assess errors in current methods
due to the quenched approximation and the finite lattice spacing in
naive lattice operators. We envision a gradual improvement of lattice
methodology to reduce these errors. Since structure functions cannot
be calculated directly on a Euclidean lattice, we consider low moments
which are related to matrix elements of local operators by the
operator product expansion.  The lattice matrix elements relevant to
moments of $F_1$, $g_1$ and $h_1$ structure functions are being
calculated on quenched $24^3\times 32$ lattices at $\beta = 6.2$ at
three quark masses, and initial results are presented below.

\section{GENERAL FORMALISM ON THE LATTICE}

In the continuum, the quark contribution to the moments of the
unpolarized $F_1$, longitudinally polarized $g_1$ and transversely
polarized $h_1$ structure functions are related to the matrix elements,
\begin{eqnarray*}
\sum_{s}\langle p s\vert{\bar\psi}\gamma_{\mu_1}D_{\mu_2}\ldots  D_{\mu_n}
\psi \vert p s\rangle &\sim& v_n(\mu),\\
\langle ps\vert{\bar\psi}\gamma_{\nu}\gamma_5D_{\mu_1} \ldots D_{\mu_n}
\psi\vert ps\rangle &\sim& a_n(\mu),\\
\langle ps\vert{\bar\psi}\sigma_{\nu\mu_1}D_{\mu_2} \ldots D_{\mu_n} 
 \psi\vert ps\rangle &\sim& t_{n}(\mu),
\end{eqnarray*}
respectively, where we have indicated on the right the scalar
coefficient in  the notation of  Refs.~\cite{DESY}.

Our methodology for calculating these operators on the lattice is
similar to G\"ockeler et al\cite{DESY}. The covariant derivative is
replaced by a lattice finite difference and the operators are
classified by representations of the surviving discrete subgroup $H_4$
of the continuum Euclidean Lorentz group $SO(4)$~\cite{Mandula}.  The
renormalization factors $Z_{\cal O}$ are calculated using a new
method~\cite{MITZ} to one loop order in lattice perturbation series to
account for a multiplicative shift of the Wilson coefficient due
replacing the $\overline{MS}$ scheme at scale $\mu$ by the lattice
spacing $a$.  Our independently derived results confirm those in the
literature~\cite{DESY,Capitani} and extend them where
necessary. For our current set of moments the results are given in
the table.

Typical elements of an irreducible representation (IR) of $H_4$ are
given in the second column, where the $\{\cdots\}$ and $[\cdots]$ brackets
imply symmetrized and anti-symmetries indices respectively. In many
cases there is more than one IR corresponding to the continuum
operator. We are exploring other representations to have a redundant
set of measurements.  The renormalization constants defined by the
formula,
\begin{displaymath}
Z_G = 1-{g_0^2\over 16\pi^2} C_F \left[\gamma_G \ln(\mu a)+B_G \right]
\end{displaymath}
are given in the third and the fourth column.

\begin{table}[htb]
\begin{center}
\begin{tabular}{|lcccr|}
\hline
&Moment && $\gamma_G$ & $B_G$ \\
\hline
\hline
\multicolumn{5}{|l|}{Spin-independent}\\
&$v_{2,a}$ & ${\cal O}_{\{14\}}$ & 16/3 & -3.160\\
&$v_{3,a}$ & ${\cal O}_{\{124\}}$ & 25/3 & -19.012\\
&$v_{4,b}$ & ${\cal O}_{\{1234\}}$ & 157/15 & -33.206 \\
\hline
\multicolumn{5}{|l|}{Spin-dependent}\\
 &$a_0$ & ${\cal O}^5_1$ & 8/3 & 4.094 \\
 &$a_1$ & ${\cal O}^5_{\{14\}}$ & 25/3 & -19.562\\
 &$a_2$ & ${\cal O}^5_{\{214\}}$ & 157/15 & -33.582 \\
\hline
\multicolumn{5}{|l|}{Tensor charge}\\
&$t_1$& ${\cal O}^{5'}_{[24]}-{\cal O}^{5'}_{[13]}$ & 2 & 16.237 \\
\hline
\end{tabular}
\end{center}
\label{tab:renorm}
\end{table}

\section{PRELIMINARY RESULTS}

On the CM-5 at MIT, we have generated 150 independent $24^3\times32$
lattices using the standard Wilson action. Dirac propagators for Wilson
fermions with $r = 1$ are being calculated by conjugate gradient
iterations with red-black preconditioning for the values of
$\kappa=0.15200, 0.15246, 0.15294$ corresponding to $m_q\approx150,
98$ and $45$ MeV. Our estimates for $\kappa_c$ and the lattice spacing
derived from $\pi$ and $\rho$ masses are in a good agreement with
those of UKQCD.

\subsection{Hadronic Sources.}
To compute the matrix elements we considered
four types of sources: point sources (P), gauge invariant
Wuppertal~\cite{Wuppertal} sources (W) and both Coulomb gauge fixed
Wuppertal (U) and Gaussian (G) smeared sources.  To determine the most
suitable form of the source, we investigated the plateau in
$\ln(G(t)/G(t+1))$ for the two point functions for the pion, rho and
nucleon sources. For the proton creation operator, we used $J_\mu(x) =
\epsilon_{abc} u^a_\mu(x)u^b_\alpha(x) \Gamma^{\alpha\beta}
d^c_\beta(x)$, with $\Gamma=C\gamma_5$ and with all quark operators
truncated to 2 upper components.

\begin{figure}[htb]
 \begin{center}
  \vskip -4ex
  \def\fpsangle{270}
  \fpshskip=-15pt
  \fpsxsize=200pt
  \fpsbox{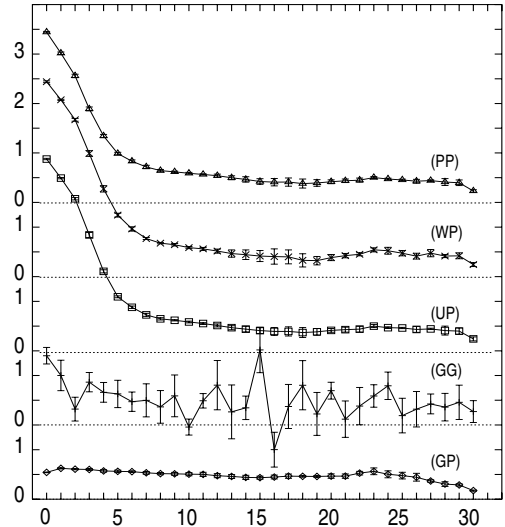}
  \vskip -3ex
  \caption{Effective mass as a function of distance 
  for the point--point (PP), Wuppertal-point (WP),
   gauge fixed Wuppertal-point (UP),
   Gaussian--Gaussian (GG)
  and  Gaussian--point (GP) nucleon correlators.}
  \label{fig:smeared}
 \end{center}
 \vskip -10ex
\end{figure}

Some comparisons are shown for the nucleon case in
Fig.~\ref{fig:smeared}. The Gaussian (G) and the two Wuppertal (W \& U)
sources were adjusted so $\sqrt{\langle x^2\rangle}\approx 6.7a$ for
each quark field, since this smearing produced the least noisy results
in all three cases.  For more localized sources the  excited states
are more prominent, whereas for less localized sources the signal
becomes noisier at large distances. 

As seen in Fig.~\ref{fig:smeared}, smearing both the sources and the
sink results in substantially noisier behavior than smearing only the
source. On the scale of the errors in Fig.~\ref{fig:smeared}, there is
no significant difference between smeared source--point sink vs. point
source--smeared sink. It is interesting to note that the gauge fixed
Wuppertal (D) and gauge invariant Wuppertal (W) sources are
essentially equivalent.

\subsection{The Tensor Charge}
We illustrate our preliminary results with
a first (low statistics) measurement of the tensor
charge~\cite{Jaffe}. (Also see results by Aoki et al\cite{Aoki}.) To
date we have analyzed $15, 10, 5$ configurations with ${\bf p} =
(0,0,0)$ for $\kappa=0.15200, 0.15246, 0.15294$ respectively and $20,
15 , 15$ configurations at ${\bf p} = (1,1,0)$ for the same kappa
values. The Gaussian smeared source is placed at $t_0=8$ and the
momentum projected point sink at $t_1=24$.

We measure three point functions in the following way. Once a gauge
field is generated, it is gauge fixed to the Coulomb gauge and a set
of quark propagators is computed.  Then for every operator the
corresponding 3-point function is constructed from the set of
propagators. We calculate separately the 2-point function, comparing
backward and forward propagators independently to verify
convergence. Finally, we use the jackknife method to estimate
statistical errors of the ratio of three- and two-point functions.
Since we need to fix the final state momentum of the nucleon before
calculating the backward propagators, the set of propagators has a
built-in final state momentum.

\begin{figure}[ht]
  \vskip -4ex
   \begin{center}
    \def\fpsangle{270}
    \fpshskip=-15pt
    \fpsxsize=300pt
    \fpsbox{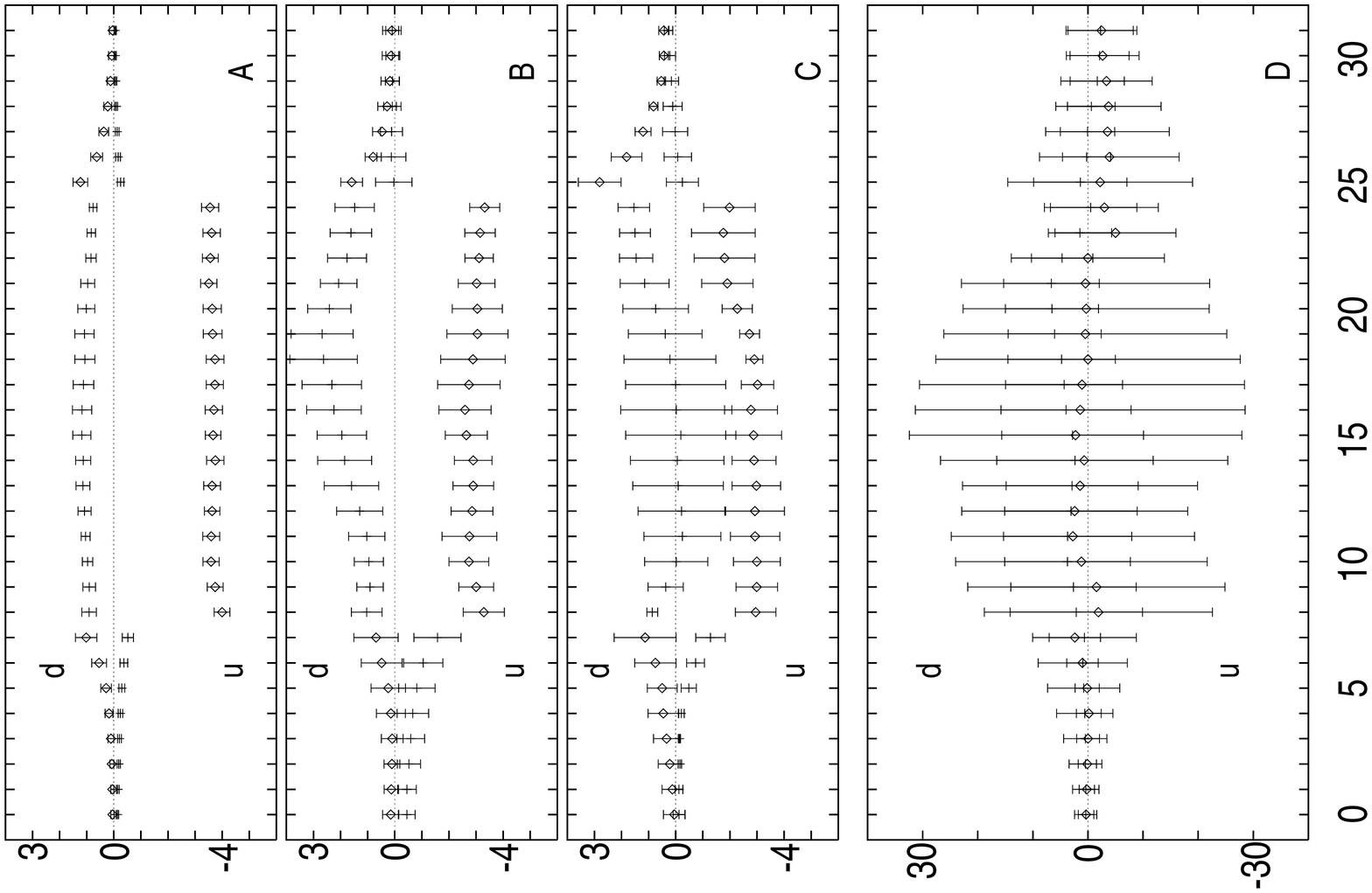}
  \vskip -3ex
  \caption{u and d contributions to $t_1$ for ${\bf p}=(0,0,0)$,
   $\kappa=0.15200$(A),$0.15246$ (B), $0.15294$ (C) and for ${\bf p}=(1,1,0)$, 
   $\kappa=0.15200$ (D).}
  \label{t0:p=0}
  \end{center}
  \vskip -10ex
 \end{figure}

The first moment $t_1$ of the transversity distribution $h_1(x)$ can
be measured using the final state at rest ${\bf p}=(0,0,0)$.
In Fig.~\ref{t0:p=0} the three upper plots show preliminary results for
$t_1$ at the three kappas we used.  Even for the low statistics (15
configurations for the smallest $\kappa$), the plateau in the signal is
clearly visible. One can also see that as $\kappa$ approaches the
chiral limit, the signal becomes noisier as expected.

Since other moments require non-zero momentum at the sink, it might be
useful to consider a single set of propagators with non-zero momenta
to measure all observables.  As a comparison, the lowest plot in
Fig.~\ref{t0:p=0}D gives the same tensor charge computed at ${\bf p} =
(1,1,0)$.  To make this a fair comparison with the ${\bf p}=(0,0,0)$
case, we have used the same set of configurations with propagators
calculated to the same precision.  It appears to have considerably
stronger fluctuations, thus raising the question as to whether the
exclusive use of non-zero momenta is an optimal  strategy.


\begin{thebibliography}{9}
\bibitem{DESY} M.~G\"ockeler, R.~Horsley, E.-M.~Ilgenfritz,
        H.~Perlt, P.~Rakow, G.~Schierholz, A.~Schiller; DESY 95--128, {\tt
         hep-lat/9508004}.
\bibitem{Mandula} J.E.~Mandula, G.~Zweig, J.~Govaerts;
       Nucl.~Phys. {\bf B228} (1983) p.~91.
\bibitem{MITZ} R.C.~Brower, S.~Huang, J.W.~Negele, A.~Pochinsky and
	B.~Schreiber, to be published (1996).
\bibitem{Capitani} G.~Beccarini, M.~Bianchi, S.~Capitani, G.~Rossi;
       Nucl.~Phys. {\bf B433} (1995) p.~351.
\bibitem{Wuppertal} S.~G\"usken, U.~L\"ow, K.H.~M\"utter, R.~Sommer,
       A.~Patel, K.~Schilling. Phys.~Lett. {\bf B227} (1989) p.~266.
\bibitem{Jaffe} R.L.~Jaffe, X.~Ji; Nucl.~Phys. {\bf B375} (1992) p.~527.
\bibitem{Aoki} S.~Aoki, M.~Doui and T.~Hatsuda; UTHEP-337,
        {\tt hep-lat/9606006}.
\end{thebibliography}
\end{document}